\documentstyle[epsf,12pt]{article}

\begin{document}

\title{One-loop counterterms for higher derivative regularized Lagrangians.}

\author{
P.I.Pronin
\thanks{E-mail:$pronin@theor.phys.msu.su$}
and K.V.Stepanyantz
\thanks{E-mail:$stepan@theor.phys.msu.su$}}

\maketitle

\begin{center}
{\em Moscow State University, physical faculty,\\
department of theoretical physics.\\
$117234$, Moscow, Russia}
\end{center}

\begin{abstract}
We explicitly calculate one-loop divergences for an arbitrary field
theory model using the higher derivative regularization and nonsingular
gauge condition. They are shown to agree with the results found in the
dimensional regularization and do not depend on the form of regularizing
term. So, the consistency of the higher derivative regularization is
proven at the one-loop level. The result for the Yang-Mills theory is
reproduced.
\end{abstract}

\sloppy

\section{Introduction.}
\hspace{\parindent}

The presence of ultraviolet divergences is the general property of
most quantum field theory models. In order to treat divergent expressions
we should made them finite by using a particular regularization scheme.

The mostly used method for making calculations is the dimensional
regularization \cite{dim}. Nevertheless, other methods sometimes can be
very useful. Really, for the investigation of chiral and
supersymmetric theories it would be better to use higher derivative
regularization \cite{slavnov}. For example, it gives the simplest proof
of chiral anomalies absence at higher loops. That is why we
believe its application for calculations to be very interesting.
However, the analysis of one-loop counterterms, using higher derivative
regularization in its original scheme presented in \cite{martin},
did not give the usual expression for $\beta$-function of the Yang-Millse
theory in the Lorentz gauge. It lead the authors to the conclusion,
that this regularization is inconsistent. Nevertheless, the discrepancy
was shown \cite{asorey} to arise due to the singular character of the
Lorentz gauge. The consistency of the original scheme with some minor
modifications was proven in \cite{bakeyev}. In particular, it was shown
how to overcome the problem of overlapping divergences \cite{book}.

Nevertheless, the higher covariant derivatives method has been mainly
used for formal arguments. We believe, that it would be interesting to
confirm the results of \cite{asorey,bakeyev} by the explicit calculations
of counterterms at least at the one-loop level. Then it can be also a
starting point for other applications of the method.

In this paper we present a way of calculating Feynman diagrams,
regularized by higher covariant derivatives. We found explicit results,
that are in agreement with the well-known expressions for the one-loop
divergences \cite{thooft,np}. They are shown to be independent on
the form of higher derivative term. So, the consistency of the method is
checked at the one-loop.

\section{Higher covariant derivatives regularization.}
\hspace{\parindent}

Let us start with the consideration of Yang-Mills theory in four
dimensional space, described by the action

\begin{equation}
S_{YM} = - \frac{1}{4} \int d^4x F_{\mu\nu}^a F^{a\mu\nu}
= \frac{1}{4g^2} tr \int d^4x F_{\mu\nu} F^{\mu\nu},
\end{equation}

\noindent
where the notations are

\begin{eqnarray}
&&F_{\mu\nu}^a=\partial_\mu A^a_\nu - \partial_\nu A^a_\mu + g f^{abc}
A_\mu^b A_\nu^c;\nonumber\\
&&F_{\mu\nu}=\partial_\mu A_\nu -\partial_\nu A_\mu +[A_\mu, A_\nu]
=-i g F_{\mu\nu}^a t^a;\nonumber\\
&&tr(t^a t^b) =\delta^{ab};\qquad [t^a, t^b] = i f^{abc} t^c;\nonumber\\
&&D_\mu = \partial_\mu + A_\mu.
\end{eqnarray}

The generating functional is given by

\begin{eqnarray}
Z[J,\eta,\bar{\eta}]=\int DA\ D\bar{c}\ Dc\
exp\Big\{i(S+S_{gf}+S_{gh} +J A+\bar{c}\eta +\bar{\eta}c)\Big\}
\end{eqnarray}

\noindent
where

\begin{equation}
J A = \int d^4x J^{\mu a} A^a_\mu
\qquad\mbox{and so on.}
\end{equation}

It is well known, that in this case the degree of divergence does not
depend on the number of loops. Nevertheless, if we add to the classical
action a term with a large degree of covariant derivatives

\begin{equation}
L \rightarrow L^\Lambda = \frac{1}{4g^2} tr \left(F^{\mu\nu} F_{\mu\nu} +
\frac{1}{\Lambda^{2n}} F^{\mu\nu} D^{2n} F_{\mu\nu} \right)
\end{equation}

\noindent
the divergences will be present only in the one-loop graphs, if $n\ge2$.
To get rid of them one usually adds Pauli-Villars (PV) fields
with large masses $M_i$. Let us consider it in more detail.

For calculations we will use the background field method \cite{dewitt}.
So, we present the gauge field as a sum of classical and quantum parts

\begin{equation}
A_\mu \rightarrow A_{\mu} + B_{\mu}
\end{equation}

\noindent
where $B_\mu$ is a quantum field and $A_\mu$ is a classical background.

As a gauge condition we choose the regularized background $\alpha$-gauge

\begin{equation}
S_{gf}^\Lambda=\frac{1}{2\alpha} \int d^4x
\left(\frac{1}{\Lambda^{2n}} \left(D^n D_\mu B^{a\mu}\right)^2
+ (D_\mu B^{a\mu})^2
\right)
\end{equation}

\noindent
where $D_\mu=D_\mu(A)$.

Then the generating functional takes the form

\begin{eqnarray}
&&Z[J]=\int DB\
exp\Big\{i(S^\Lambda(A+B)+S_{gf}^\Lambda +J B)\Big\}\nonumber\\
&&\qquad\qquad\qquad\qquad\qquad\qquad
\times det(D^2)det^{1/2}\Big(1+\frac{D^{2n}}{\Lambda^{2n}}\Big).
\end{eqnarray}

In order to regularize the remaining one-loop divergences we add
PV-fields, so that

\begin{eqnarray}\label{zz}
&&Z[J]=\int DB\
exp\Big\{i(S^\Lambda(A+B)+S_{gf}^\Lambda+ J B)\Big\}\nonumber\\
&&\qquad\qquad\qquad\qquad\qquad\qquad
\times det G \prod\limits_i det^{-a_i/2}D_{M_i} det^{b_i}G_{m_i}
\end{eqnarray}

\noindent
where we introduced the notations

\begin{eqnarray}\label{det}
&&D^\Lambda{}^{\mu\nu}_{ab}
= \frac{\delta^2 S^\Lambda}{\delta A_\mu^a \delta A_\nu^b};\nonumber\\
&&D^\Lambda{}_{M_i}{}^{\mu\nu}_{ab} = \frac{1}{\Lambda^{2n}}
\frac{\delta^2 S^\Lambda_0}{\delta A_\mu^a \delta A_\nu^b}
- g_{\mu\nu} \delta^{ab} M_i^2;\nonumber\\
&& G^\Lambda{}^{ab} =
\left(\Big(1+\frac{D^{2n}}{\Lambda^{2n}}\Big)^{1/2} G\right)^{ab}
;\nonumber\\
&& G^\Lambda{}_{m_i}{}^{ab} =
\left(\Big(1+\frac{D^{2n}}{\Lambda^{2n}}\Big)^{1/2}
(G^\Lambda{}- m_i^2)\right)^{ab}.
\end{eqnarray}

\noindent
Here $G = D^2$ is a ghost determinant without regularization and
$S^\Lambda_0/\Lambda^{2n}$ is a term with higher derivatives.

The coefficients $a_i$ should satisfy the following conditions

\begin{eqnarray}
&&\sum_i a_i = -1;\ \qquad \sum_i b_i = -1; \nonumber\\
&&\sum_i a_i M_i^2 = 0\qquad \sum_i b_i m_i^2 =0.
\end{eqnarray}

Let us note, that this conditions provide the complete cancellation of
higher derivatives determinants for the ghost fields in (\ref{zz}), that
can be, therefore, omitted.

\section{One-loop diagrams calculation.}
\hspace{\parindent}

At the one-loop level using (\ref{zz}) we easily obtain

\begin{equation}\label{efac}
\Gamma^{(1)}_\infty =
\frac{i}{2} \mbox{tr}
\ln \Big(D^\Lambda \prod\limits_i (D_{M_i}^\Lambda)^{a_i}\Big)
-i\ \mbox{tr} \ln \Big(G \prod\limits_i (G_{m_i})^{b_i}\Big).
\end{equation}

In this section we explicitly find the logarithmically divergent part of
(\ref{efac}) and show, that it is in agreement with general results
\cite{thooft,np}. We do not want to restrict ourselves by the analysis of
the pure Yang-Mills theory and consider the most general theory with the
action $S(\phi^I)$ in the flat space. Then equation (\ref{efac}) is valid if

\begin{eqnarray}
&&D^\Lambda{}_I{}^J=
\frac{\delta^2 S^\Lambda}{\delta \phi^I \delta \phi_J}
;\nonumber\\
&&D^\Lambda{}_{M_i}{}_I{}^J = \frac{1}{\Lambda^{2n}}
\frac{\delta^2 S^\Lambda_0}{\delta \phi^I \delta \phi_J}
 - \delta_I{}^J M_i^2;\nonumber\\
&& G_{m_i}{}_i{}^j = G_i{}^j - \delta_i{}^j m_i^2,
\end{eqnarray}

\noindent
where we take into account the cancellation of ghost determinants with
higher derivatives. (Here capital letters denote $\phi$ indexes, while
small letters - the ghost ones).

Let us proceed to the construction of the diagram technique. Extracting the
regularizing term $S^\Lambda_0$, we rewrite the action as

\begin{equation}\label{sps}
S^\Lambda = S + \frac{1}{\Lambda^{2n}} S^\Lambda_0,
\end{equation}

\noindent
so that

\begin{equation}
D^\Lambda = D + \frac{1}{\Lambda^{2n}} D^\Lambda_0.
\end{equation}

The most general form of differential operators $D$ and $D^\Lambda_0$
is

\begin{eqnarray}\label{nonminimal}
&&
\vphantom{\frac{1}{2}}
D_{I}{}^{J} =
K^{\mu_1\mu_2\ldots \mu_{L}}{}_{I}{}^{J}
\ \nabla_{\mu_1} \nabla_{\mu_2}\ldots
\nabla_{\mu_{L}}
+\ S^{\mu_1\mu_2\ldots \mu_{L-1}}{}_{I}{}^{J}
\ \nabla_{\mu_1} \nabla_{\mu_2}\ldots
\nabla_{\mu_{L-1}}
\nonumber\\
&&
\vphantom{\frac{1}{2}}
+\ W^{\mu_1\mu_2\ldots \mu_{L-2}}{}_{I}{}^{J}
\ \nabla_{\mu_1} \nabla_{\mu_2}\ldots
\nabla_{\mu_{L-2}}
+\ N^{\mu_1\mu_2\ldots \mu_{L-3}}{}_{I}{}^{J}
\ \nabla_{\mu_1} \nabla_{\mu_2}\ldots
\nabla_{\mu_{L-3}}
\nonumber\\
&&
\vphantom{\frac{1}{2}}
+\ M^{\mu_1\mu_2\ldots \mu_{L-4}}{}_{I}{}^{J}
\ \nabla_{\mu_1} \nabla_{\mu_2}\ldots
\nabla_{\mu_{L-4}} +
\ldots,
\end{eqnarray}

\noindent
where $\nabla_\mu$ is a covariant derivative

\begin{eqnarray}
\nabla_\mu \Phi_I = \partial_\mu \Phi_I + \omega_{\mu I}{}^J \Phi_J,
\end{eqnarray}

\noindent
and $\omega_{\mu I}{}^J$ is a connection on the principle bundle.
(In this paper we consider only theories in the flat space.)

Higher derivatives regularization assumes that $L^\Lambda > L+4$,
where $L^\Lambda$ is the order of operators $D^\Lambda$ and $D^\Lambda_0$
and $L$ is the order of operator $D$. So, according to the general analysis
given in \cite{np}, remaining one-loop divergences (before taking the limit
$\Lambda \rightarrow 0$) are defined only by the
form of high derivative regularizing term $D^\Lambda_0$. Therefore, they
are completely canceled by the contribution of PV-fields and the final
expression is finite.

Now we are going to present (\ref{efac}) as a sum of one-loop diagrams.
For this purpose we extract the terms with the largest number of
derivatives from $D$ and $D^\Lambda_0$ and make the expansion

\begin{eqnarray}
\frac{i}{2}\ \mbox{tr}\ \ln\ D^\Lambda =
\frac{i}{2}\ \mbox{tr}\ \ln\ \left((K^\Lambda\partial) + V^\Lambda \right)
= \frac{i}{2}\ \mbox{tr} \sum_{k=1}^\infty \frac{1}{k}
\left(-\ \frac{1}{(K^\Lambda\partial)} V^\Lambda\right)^k.
\end{eqnarray}

\noindent
where

\begin{eqnarray}
&&K^\Lambda = \frac{1}{\Lambda^{2n}} K^\Lambda_0 + K\nonumber\\
&&V^\Lambda \equiv D^\Lambda -(K^\Lambda\partial)
= \frac{1}{\Lambda^{2n}} V^\Lambda_0 + V
\end{eqnarray}

\noindent
and $K^\Lambda_0$ comes from the expansion of higher derivative term
$S_0$ in (\ref{sps}).

For PV-determinants the corresponding expansion is

\begin{eqnarray}
&&\frac{i}{2}\ \mbox{tr}\ \ln\ D^\Lambda_M =
\frac{i}{2}\ \mbox{tr}\ \ln\
\left(\Big((K^\Lambda_0\partial)/\Lambda^{2n} - M^2\Big) + V^\Lambda \right)
\nonumber\\
&&\qquad\qquad\qquad\qquad\qquad\qquad
= \frac{i}{2}\ \mbox{tr} \sum_{k=1}^\infty \frac{1}{k}
\left(-\ \frac{1}{(K^\Lambda_0\partial)/\Lambda^{2n} - M^2}
V^\Lambda\right)^k.
\end{eqnarray}

We see, that for each diagram, containing quantum fields, there
are graphs with corresponding PV-fields. As we mentioned above,
their sum is finite.

The result of integration over a loop momentum depends on a way of taking
the limit $M \rightarrow \infty, \Lambda \rightarrow \infty$.
Nevertheless, it does not affect renormgroup functions \cite{martin}.
So, we are able to choose for calculations the simplest prescription
$M_i/\Lambda = \theta_i=const$.

Here we should note, that using of the higher derivative regularization
(unlike dimensional regularization) leads to the nontrivial contribution
of quadratically divergent graphs to the effective action. In this paper
we do not intend to find it and restrict ourselves only by checking the
agreement of logarithmic divergences with the result obtained in the
dimensional regularization.

Let us consider expressions for Feynman diagrams with the PV-contribution

\begin{eqnarray}
\frac{i}{2}\ \mbox{tr} \sum_{k=1}^\infty \frac{1}{k}
\left\{
\left(-\ \frac{1}{(K^\Lambda\partial)} V^\Lambda\right)^k
+ \sum\limits_i
a_i\left(-\ \frac{1}{(K^\Lambda_0\partial)/\Lambda^{2n}
- \theta_i^2\Lambda^2}
V^\Lambda\right)^k
\right\}
\end{eqnarray}

For any finite $\Lambda$ they are also UV-finite. Let us first extract
the terms, that diverge at $\Lambda = \infty$. Then we turn to the
momentum representation and expand the integrand in powers of external
momentum, keeping only logarithmical divergences. \footnote{Of course,
they can arise from more than logarithmically divergent terms; for the
detailed discussion see \cite{np}.} Their total contribution is UV-finite
if $\Lambda \ne \infty$. Nevertheless, there are IR-divergences, that can
be regularized by cutting

\begin{equation}
\int\limits_0^\infty dk \rightarrow \int\limits_m^\infty dk
\end{equation}

\noindent
Due to the logarithmic divergence and scale invariance the results of
calculations will be proportional to $\ln (\Lambda/m)$. So, instead of
finding UV-assimptotics, we can calculate IR one, that is much easier.
(We should, of course, keep terms, that diverge in the limit
$m \rightarrow 0$). Really, in this limit (if $n \ge 2$) we should omit
regularizing terms, coming from $S^\Lambda_0$, because they are much
smaller than the original action, while PV-diagrams contains

\begin{equation}
\int d^4k \frac{1}{M^4}
\end{equation}

\noindent
and do not contribute to terms of $\ln m$ order.

So, we should omit  regularizing term, PV-fields and find the assimptotics
of the remaining logarithmically divergent integrals

\begin{equation}
\left(\int d^4k \frac{1}{k^4}\right)_{reg} =
2i\pi^2 \int\limits_m dk \frac{1}{k} = - 2i\pi^2 \ln m
\end{equation}

\noindent
where $i$ comes from $k_0=i k_4$. (Momentum cutting should be performed
only in the Euclidean space).

Taking into account that the result is proportional to $\ln (\Lambda/m)$,
we obtain the following prescription for obtaining logarithmically
divergent counterterms: omit the higher derivative term, PV-diagrams,
expand the remaining expression in powers of the loop momentum and,
finally, make a substitution

\begin{equation}
\int d^4k \frac{1}{k^4} \quad\mbox{на}\quad 2i\pi^2 \ln\Lambda.
\end{equation}

Of course, it does not mean, that terms with higher derivatives or
PV-fields are insignificant. They provide the UV-finiteness of the
whole sum, that was very important for calculations. The independence of
the result on the particular form of higher derivative term is quite
natural and should be expected originally.

The above prescription is very similar to the one found in \cite{hep1}
for the dimensional regularization. The only difference is that the
substitution in that case is

\begin{equation}
\int d^dk \frac{1}{k^4} \rightarrow - \frac{2i\pi^2}{d-4}
\equiv \frac{2i\pi^2}{\epsilon}.
\end{equation}

This prescription was used in \cite{np} for the explicit calculation
of the effective action. It means, that everything has been already done
and we are able to write down the logarithmically divergent terms for
an arbitrary Lagrangian. For example, if

\begin{equation}
D_{i}{}^{j} = \delta_{i}{}^{j} \Box + S^\mu{}_{i}{}^{j} \nabla_\mu +
W_{i}{}^{j}.
\end{equation}

\noindent
then

\begin{eqnarray}\label{secondorder}
&&\Gamma^{(1)}_{\infty,\ln} = \ln \Lambda^2 \frac{1}{32\pi^2}
 \mbox{tr} \int d^4 x
\left(\frac{1}{12} Y_{\mu\nu} Y^{\mu\nu}
+ \frac{1}{2} X^2\right)
\end{eqnarray}

\noindent
where

\begin{eqnarray}
&&Y_{\mu\nu} =
\frac{1}{2}\nabla_\mu S_\nu - \frac{1}{2}\nabla_\mu S_\nu
+ \frac{1}{4} S_\mu S_\nu - \frac{1}{4} S_\nu S_\mu
+ F_{\mu\nu},\nonumber\\
&&X = W - \frac{1}{2} \nabla_\mu S^\mu - \frac{1}{4} S_\mu S^\mu
\end{eqnarray}

For more complicated examples we should substitute $1/\epsilon$ by
$\ln \Lambda$ in the general expressions, found in \cite{np}.

So, the logarithmically divergent part of the one-loop effective action
is completely the same as for the dimensional regularization and one-loop
consistency of the higher derivative regularization is checked explicitly
at the one-loop level.

In the end we should note, that performing the above calculations we
assume $\alpha\ne 0$. Therefore, our results are in agreement with
\cite{asorey}.

Now we can easily find the logarithmically divergent counterterms for
the Yang-Millse theory. The second variation of the action in the
$\alpha$-gauge is the following differential operator

\begin{equation}\label{vector}
D_\alpha{}^\beta = \Box\ \delta_{\alpha}{}^{\beta} - \frac{\alpha-1}{\alpha}
\ \nabla_\alpha \nabla^\beta + 2 F_{\alpha}{}^\beta.
\end{equation}

The calculations performed by the algorithms obtained in \cite{np}
gives the result

\begin{equation}
\Gamma^{(1)}_{\infty,\ln} = \frac{1}{32\pi^2} \ln \Lambda^2
\frac{11}{3} C_1 F_{\mu\nu}^a F^{\mu\nu a}.
\end{equation}

\noindent
that does not depend on the gauge parameter.

\section{Conclusion.}
\hspace{\parindent}

In this paper we explicitly check at the one-loop level, that the
regularization by higher covariant derivatives is a consistent method
applicable for calculations of quantum corrections. For an arbitrary
Lagrangian we found that the counterterms are the same as the ones for
dimensional regularization up to the substitution
$1/\epsilon \rightarrow \ln \Lambda$ (except for the singular gauge
conditions) and does not depend on the particular form of higher
derivative term.

It is also worth mentioning, that higher derivative regularization allows
to calculate quadratic divergences, although we do not consider them in the
present paper.


\vspace{1cm}

\noindent
{\Large\bf Acknowledgments.}

\vspace{0.7cm}

We are very grateful to Prof. Slavnov A.A. (Steklov Mathematical
Institute) for the numerous discussions and continuous attention
to our work and especially like to thank Asadov V.V. for the financial
support.


\end{document}